\documentclass{elsart}
\usepackage{amsfonts}
\usepackage{amssymb}
\usepackage{graphicx}
\usepackage{dcolumn}
\usepackage{bm}

\newtheorem{theorem}{Theorem}
\newtheorem{definition}{Definition}

\newtheorem{example}{Example}

\def\0{\mbox{\bf 0}}
\def\BibTeX{{\rm B\kern-.05em{\sc i\kern-.025em b}\kern-.08em
    T\kern-.1667em\lower.7ex\hbox{E}\kern-.125emX}}
\input{tcilatex}

\begin{document}


\begin{frontmatter}

\title{Forbidden ordinal patterns in higher dimensional dynamics}

\author{Jos\'{e} M. Amig\'{o}$^1$ and Matthew B. Kennel$^2$}


\address{
$^1$Centro de Investigaci\'{o}n Operativa, \\
Universidad Miguel Hern\'{a}ndez. 03202 Elche, Spain \\
$^2$Institute for Nonlinear Science, University of California, \\
San Diego. La Jolla, CA 92093-0402, USA \\
%
%
\mbox{E-mails}: {\tt jm.amigo@umh.es, mkennel@ucsd.edu}}

\begin{abstract}
Forbidden ordinal patterns are ordinal patterns (or `rank blocks')
that cannot appear in the orbits generated by a map taking values on
a linearly ordered space, in which case we say that the map has
forbidden patterns. Once a map has a forbidden pattern of a given
length $L_{0}$, it has forbidden patterns of any length $L\ge L_{0}$
and their number grows superexponentially with $L$. Using recent
results on topological permutation entropy, we study in this paper
the existence and some basic properties of forbidden ordinal
patterns for self maps on $n$-dimensional intervals. Our most
applicable conclusion is that expansive interval maps with finite
topological entropy have necessarily forbidden patterns, although we
conjecture that this is also the case under more general conditions.
The theoretical results are nicely illustrated for $n=2$ both using
the naive counting estimator for forbidden patterns and Chao's
estimator for the number of classes in a population. The robustness
of forbidden ordinal patterns against observational white noise is
also illustrated.
\end{abstract}


\end{frontmatter}


\section{\noindent Introduction}

\noindent Ordinal patterns of length $L$ describe the relations of
`smaller'\ or `larger'\ among $L$ consecutive points of a deterministically
or randomly generated sequence in a linearly ordered space. Ordinal patterns
are the main ingredient of \textit{permutation entropy}, a concept
introduced both in metric and topological versions by Bandt, Keller and
Pompe \cite{Bandt}, that were shown to coincide with their standard
counterparts for piecewise monotone one-dimensional interval maps. Later on,
the concepts of metric and topological permutation entropies were
generalized to $n$-dimensional interval maps in \cite{Amigo} and \cite%
{Amigo3}, respectively, while preserving the main results of \cite{Bandt}
although under different assumptions: ergodicity for the metric entropy and
expansiveness for the topological entropy. Both generalizations parallel
Kolmogorov's construction of entropy in dynamical systems in that they
coarse-grain the state space with partitions, apply the corresponding
definition of entropy to the resulting symbolic dynamics and, lastly, take
ever finer partitions. But this time the partitions used are product,
uniform partitions (the original, one-dimensional versions even dispense
thoroughly with partitions), making possible, albeit computationally
demanding, the numerical estimation of metric and topological entropy.
Moreover, ordinal patterns allow a unified and conceptually simple approach
both to metric and topological entropy, at variance with the standard
approach.

In this paper, that can be considered a second part of \cite{Amigo3}, we
deal only with the topological permutation entropy. Having shown in \cite%
{Amigo3} that this concept converges to topological entropy for $n$%
-dimensional \textit{expansive} interval maps and illustrated how it can be
used as estimator, we focus our attention now on some interesting
consequences of order in dynamical systems.

First of all, it turns out that the orbits of \textit{continuous}, $n$%
-dimensional interval maps with finite topological \textit{permutation
entropy} have always forbidden patterns, i.e., ordinal patterns that cannot
occur in the orbits of the map, in contrast with (unconstrained) random time
series, in which any ordinal patterns appears with probability $1$. As a
more practical result, it follows that the same happens to expansive maps
with finite topological entropy. Furthermore, forbidden patterns proliferate
superexponentially with length, the exact details being controlled by the
topological permutation entropy. The existence and growth rate of forbidden
patterns was already considered in \cite{Amigo2,Amigo5,Amigo4} but in a
rather restrictive setting, namely, for piecewise monotone maps on
one-dimensional intervals only (i.e., using the original definition of
topological permutation entropy and results of \cite{Bandt}). In the present
paper, we go higher dimensional by using the definitions and results of \cite%
{Amigo3}.

Secondly, forbidden patterns are, in general, not invariant under
isomorphism (or conjugacy) between dynamical systems unless the isomorphism
is order-preserving, i.e., it is an order-isomophism. This allows to further
subdivide isomorphic systems according to their forbidden patterns, thus
opening the door to more restrictive definitions of equivalence among maps
of $n$-dimensional intervals.

Last but not least, forbidden patterns are robust against observational
noise on account of being defined by inequalities. Robustness was shown in
\cite{Amigo5} to be instrumental for practical applications, specifically in
scalar time series analysis. Indeed, forbidden patterns can discriminate
deterministic from random time series when the noise is white, even if the
noise level is so high that any trace of determinism is washed out in the
return map graph. The case with colored noise is currently under
investigation.

This paper is organized as follows. Sect. 2 explains the basic conceptual
and notational framework. Sect. 3 and 4 are devoted to the study of
forbidden ordinal patterns in higher dimensional dynamics; the former
contains the theoretical core and the latter deals with more practical
issues, like the structure of forbidden patterns and their robustness
against noise. Sect. 5 illustrates the theoretical sections with numerical
evidence of forbidden patterns for Arnold's cat map and H\'{e}non's map,
both using the naive counting estimator and Chao's estimator for the number
of classes in a population, which is quite popular in mathematical biology
for estimating the number of species in ecological systems. The robustness
of forbidden patterns against observational white noise is also addressed.

\section{\noindent Preliminaries and previous work}

\subsection{Topological permutation entropy of information sources}

A finite-state (resp. finite-alphabet) \textit{information source} with
states (resp. alphabet) $A=\{a_{1},a_{2},...,a_{\left\vert A\right\vert }\}$
is a \textit{stationary} stochastic process $\mathbf{S}=(S_{n})_{n\in
\mathbb{N}_{0}}$ on a probability space $(\Omega ,\mathcal{F},\mu )$. Here $%
\mathbb{N}_{0}:=\{0\}\cup \mathbb{N}$, $\Omega $ is a non-empty set, $%
\mathcal{F}$ is a sigma-algebra of subsets of $\Omega $, $\mu $ is a
probability measure on the measurable space $(\Omega ,\mathcal{F})$, and $%
S_{n}:\Omega \rightarrow A$ are random variables. Such random processes
provide models for physical information sources that must be turned on at
some time. Observe that the possible outputs (realizations, messages,...) $%
\mathbf{S}(\omega )=(s_{n})_{n\in \mathbb{N}_{0}}$, $\omega \in \Omega $, of
the process $\mathbf{S}$ are points in the \textit{sequence space}%
\begin{equation}
A^{\mathbb{N}_{0}}=\{\mathbb{\alpha }=(\mathbb{\alpha }_{n})_{n\in \mathbb{N}%
_{0}}=(\alpha _{0},\alpha _{1},...):\mathbb{\alpha }_{n}\in A\}.
\label{sequence}
\end{equation}

A relation $\leq $ on a set $X$ is said to be a total order (or $X$ to be a
totally ordered set) if $\leq $ is reflexive, antisymmetric and transitive,
and moreover all elements of $X$ are comparable. As usual, $x<y$ ($x,y\in X$%
) means henceforth $x\leq y$ and $x\neq y$. The product of linearly ordered
sets, $(X_{1},\leq )$, $(X_{2},\leq )$,..., $(X_{n},\leq )$, is also
linearly ordered via the product (also called lexicographical or dictionary)
order: if $(x_{1},x_{2},...,x_{n})\neq (y_{1},y_{2},...,y_{n})$, then $%
(x_{1},x_{2},...,x_{n})<(y_{1},y_{2},...,y_{n})$ if (\textit{i}) $%
x_{1}<y_{1} $, or (\textit{ii}) $x_{i}=y_{i}$ for $i=1,...,k$, where $1\leq
k\leq n-1$, and $x_{k+1}<y_{k+1}$; other conventions are of course possible.
The product order generalizes straighforwardly to \textquotedblleft infinite
products\textquotedblright\ (i.e., sequences spaces).

Suppose now that the alphabet $A$ of the information source $\mathbf{S}$ is
endowed with a total ordering $\leq $, and let $\mathcal{S}_{L}$ denote the
set of permutations on $\{0,1,...,L-1\}$. If $\pi \in \mathcal{S}_{L}$ and $%
0\mapsto \pi _{0}$, ..., $L-1\mapsto \pi _{L-1}$, then we write $\pi
=\left\langle \pi _{0},...,\pi _{L-1}\right\rangle $. Given the output $%
(s_{n})_{n\in \mathbb{N}_{0}}$ of $\mathbf{S}$, we say that a length-$L$ word%
\textit{\ }$s_{n}^{n+L-1}=s_{n},s_{n+1},...,s_{n+L-1}$ defines the \textit{%
ordinal }($L$-)\textit{pattern} $\pi \in \mathcal{S}_{L}$\ if\footnote{%
In the references, \textquotedblleft ordinal patterns\textquotedblright\ are
called \textquotedblleft order patterns\textquotedblright\ and written
between rectangular (instead angular) parentheses.}
\[
s_{n+\pi _{0}}\prec s_{n+\pi _{1}}\prec ...\prec s_{n+\pi _{L-1}},
\]%
where, for definiteness, given $s_{i},s_{j}\in A$ and $i,j\in \mathbb{N}_{0}$
with $i\neq j$,%
\[
s_{i}\prec s_{j}\;\Leftrightarrow \;\left\{
\begin{array}{cc}
s_{i}<s_{j} &  \\
\text{\mbox{or}} &  \\
i<j & \text{\mbox{if} }s_{i}=s_{j}.%
\end{array}%
\right. .
\]

For example, suppose that $\mathbf{S}$ is a source over the alphabet $%
A=\{1,2,3\}$ ordered by size, and that we observe the output $%
s_{0}^{2}=3,1,1 $. Then, the word $s_{0}^{2}$ defines the ordinal pattern $%
\left\langle 1,2,0\right\rangle $. Formally one can associate with $\mathbf{S%
}$ a (non-stationary) random process $\mathbf{R}=(R_{n})_{n\in \mathbb{N}_{0}%
\text{ }}$, $R_{n}:\Omega \rightarrow \mathbb{N}$, via $R_{n}=\left\vert
\{S_{i},0\leq i\leq n:S_{i}\leq S_{n}\}\right\vert $ ($\left\vert \cdot
\right\vert $ denotes cardinality), whose outputs (`ranks') are in a
one-to-one relation with the ordinal patterns defined by the outputs of $%
\mathbf{S}$.

The metric and topological \textit{permutation} entropies of an information
source are defined analogously to the metric and topological entropies, but
using \textit{ordinal patterns} instead of words. In particular, if $%
S_{0}^{L-1}$ is shorthand for the block of random variables $%
S_{0},...,S_{L-1}$ and $N(\mathbf{S},L)$ is the number of \textit{allowed
ordinal }$L$-\textit{patterns} that $\mathbf{S}$ can output (or,
equivalently, the number of length-$L$ words of the form $%
r_{0}^{L-1}=r_{0},...,r_{L-1}$ that can be observed in the messages of $%
\mathbf{R}$), we have:

\begin{defn}
\label{Top1and2}The \textit{topological permutation entropy of order} $L$ of
$\mathbf{S}$ is defined as%
\begin{equation}
H_{top}^{\ast }(S_{0}^{L-1})=\frac{1}{L-1}\log N(\mathbf{S},L),  \label{top1}
\end{equation}%
and the \textit{topological permutation entropy} of $\mathbf{S}$ as%
\begin{equation}
h_{top}^{\ast }(\mathbf{S})=\lim_{L\rightarrow \infty }\sup H_{top}^{\ast
}(S_{0}^{L-1}).  \label{top2}
\end{equation}
\end{defn}

The normalization factor $1/(L-1)$ in (\ref{top1}) instead of $1/L$, is due
to the fact that single letters do not define any ordinal pattern (of
course, the choice $1/L$ leads to the same limit when $L\rightarrow \infty $%
). The logarithm in (\ref{top1}) can be taken to any base $>1,$ the most
usual bases being $2$ ($h_{top}^{\ast }$ in units of bits per symbol) and $e$
($h_{top}^{\ast }$ in units of nits per symbol). For convenience, we will
use Neperian logarithms.

\subsection{Topological permutation entropy of maps}

For our needs it is sufficient to restrict the definition of topological
permutation entropy of maps to interval maps. Let $I$\ be a finite interval
of $\mathbb{R}^{q}$ and $f:I\rightarrow I$ a $\mu $-preserving map, with $%
\mu $ being a probability measure on $I$ endowed with the Borel
sigma-algebra $\mathcal{B}$. In order to define next the topological
permutation entropy of $f$, we consider first a special coarse-graining made
out of a product partition
\[
\iota =\prod_{k=1}^{q}\{I_{1,k},\ldots ,I_{N_{k},k}\}
\]%
of $I$ into $N:=N_{1}\cdot ...\cdot N_{q}$ subintervals of lengths $\Delta
_{j,k}$, $1\leq j\leq N_{k}$, in each coordinate $k$, defining the \textit{%
norm }$\left\Vert \iota \right\Vert =\max_{j,k}\Delta _{j,k}$ of the
partition $\iota $ (other definitions are also possible). For definiteness,
the intervals are \textit{lexicographically} ordered in each dimension,
i.e., points in $I_{j,k}$ are smaller than points in $I_{j+1,k}$ and, for
the multiple dimensions, a \textit{lexicographic} order is defined, $%
I_{j,k}<I_{j,k+1}$, so there is an order relation between all the $N$
partition elements, and we can enumerate them with a single index $i\in
\{1,...,N\}$:%
\[
\iota =\{I_{i}:1\leq i\leq N\}\text{, \ }I_{i}<I_{i+1}.
\]

Next define a collection of \textit{simple observations} $\mathbf{S}^{\iota
}=(S_{n}^{\iota })_{n\in \mathbb{N}_{0}}$ with respect to $f$ with precision
$\left\Vert \iota \right\Vert $:
\[
S_{n}^{\iota }(x)=i\;\;\text{\mbox{if}}\;\;f^{n}(x)\in I_{i},n=0,1,\ldots
\]%
Then $\mathbf{S}^{\iota }$ is a stationary $N$-state random process or,
equivalently, an information source on $(I,\mathcal{B},\mu )$ with finite
alphabet $A^{\iota }=\{1,...,N\}$. In a dynamical setting, $\mathbf{S}%
^{\iota }$ is called the symbolic dynamic with respect to the coarse
graining $\iota $.

\begin{defn}
\label{Top(f)}The topological permutation entropy of $f$ is defined as
\begin{equation}
h_{top}^{\ast }(f)=\lim_{\left\Vert \iota \right\Vert \rightarrow
0}h_{top}^{\ast }(\mathbf{S}^{\iota }).  \label{DefB}
\end{equation}
\end{defn}

Note that the limit (\ref{DefB}) exists since $h_{top}^{\ast }(\mathbf{S}%
^{\iota })$ is non-decreasing with ever finer partitions $\iota $. Moreover,
this limit can be shown not to depend on the particular partition $\iota $,
so that $\iota $ may be taken to be uniform (i.e., a `box partition')
without restriction. This being the case, we will consider in the sequel
only box partitions.

If $f$ is continuous, $h_{top}^{\ast }(f)$ is an upper bound of the
topological entropy of $f$, $h_{top}(f)$ \cite{Amigo3}:%
\[
h_{top}(f)\leq h_{top}^{\ast }(f).
\]%
One of the main interests of $h_{top}^{\ast }(f)$ is that, under an
additional hypothesis on $f$, it can be shown to coincide with the
topological entropy of $f$, $h_{top}(f)$, thus providing eventually an
estimator of it \cite{Amigo3}.

Indeed, let $(X,d)$ be a \textit{compact} metric space, $d$ denoting a
metric on $X$. A homeomorphism (correspondingly, a continuous map) $%
f:X\rightarrow X$ is said to be \textit{expansive} if there exists $\delta
>0 $ such that $d(f^{n}(x),f^{n}(y))\leq \delta $ for all $n\in \mathbb{Z}$
(correspondingly, $n\in \mathbb{N}_{0}$) implies $x=y$. We will call $\delta
$ the \textit{expansiveness constant} of $f$. Intuitively, the orbits of an
expansive map $f$ can be resolved to any desired precision by taking $n$
sufficiently large. Standard examples of expansive maps include expanding
maps on the circle, topological Markov chains, hyperbolic toral
automorphisms and shift transformations on sequence spaces \cite{Katok}.

\begin{theorem}
\label{main}\cite{Amigo3} If $I\subset \mathbb{R}^{q}$ is a compact interval
and $f:I\rightarrow I$ an expansive map, then
\[
h_{top}^{\ast }(f)=h_{top}(f).
\]
\end{theorem}

This theorem holds also true for restrictions of expansive maps on open or
half-open subintervals.

\section{Forbidden ordinal patterns}

Let $f:X\rightarrow X$ be a map, where $X$ is endowed with a linear order $%
\leq $. We say that $x\in X$ defines the ordinal ($L$-)pattern $\pi
=\left\langle \pi _{0},\pi _{1},...,\pi _{L-1}\right\rangle \in \mathcal{S}%
_{L}$ if
\begin{equation}
f^{\pi _{0}}(x)<f^{\pi _{1}}(x)<...<f^{\pi _{L-1}}(x)  \label{patt}
\end{equation}%
(where $f^{0}(x)\equiv x$ and $f^{n}(x)\equiv f(f^{n-1}(x))$).
Alternatively, we say that $\pi $ is an \textit{allowed} (\textit{ordinal})
\textit{pattern} of $f$, or that $\pi $ is allowed, in which case%
\[
P_{\pi }\equiv \{x\in I:x\text{ \mbox{defines} }\pi \in \mathcal{S}%
_{L}\}\neq \varnothing \text{.}
\]%
Furthermore, $\pi ^{\prime }\in \mathcal{S}_{L}$ is said to be a \textit{%
forbidden} (\textit{ordinal}) \textit{pattern} for $f$, or just to be
forbidden, if there exists no $x\in I$ defining $\pi ^{\prime }$, i.e., $%
P_{\pi ^{\prime }}=\varnothing $. If $f$ is continuous (as we will consider
below), then the sets $P_{\pi }\subset X$ are open for all $\pi \in \mathcal{%
S}_{L}$, $L\geq 2.$

Henceforth we consider self maps $f$ on intervals $I\subset \mathbb{R}^{q}$
endowed with lexicographical order. In this way, the ordinal patterns of $f$
will coincide with the ordinal patterns of the the corresponding symbolic
dynamic $\mathbf{S}^{\iota }$ with respect to the box partition $\iota
=\{I_{i}\}_{1\leq i\leq N}$ in the limit $\left\Vert \iota \right\Vert
\rightarrow 0$. Indeed, if $x\in I$ defines the pattern $\pi \in \mathcal{S}%
_{L}$, the only way that the length-$L$ word $x,f(x),...,f^{L-1}(x)$ does
not define $\pi $ when observed with the precision set by the
coarse-graining $\iota $ is that at least two letters, say $f^{i}(x)$ and $%
f^{j}(x)$, $0\leq i<j\leq L-1$, fall in the same box $I_{i_{0}}\in \iota $
(since then we cannot discern the order relation between both letters). But
all these exceptions will disappear in the limit $\left\Vert \iota
\right\Vert \rightarrow 0$.

The existence of forbidden patterns is obvious in some trivial cases, e.g.,
when $f$ is periodic. Also, one can set up `by hand' functions with no
forbidden patterns (in particular, discontinuous one-dimensional interval
maps with infinite many monotony segments). Let us mention in passing that
the forbidden patterns of a one-dimensional maps can easily be exposed via
the graphs of the map and its iterates.

\begin{figure}[tbp]
\centerline{\includegraphics[width=0.7%
\columnwidth]{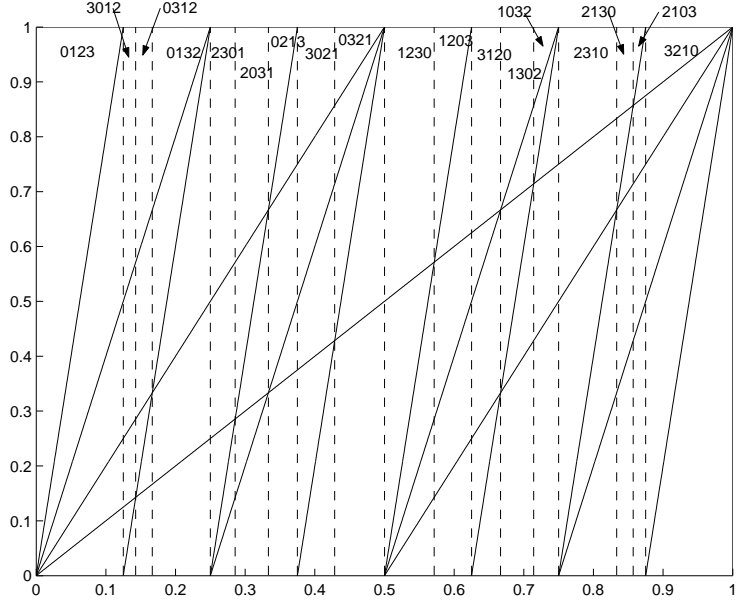}} \label{fig:shiftmap}
\caption{ The $18$ allowed 4-patterns of the shift map $S_{2}:x\mapsto 2x (%
\func{mod}1)$. For clarity, the allowed patterns have been written without
angular parentheses nor separating commas. Note that the allowed patterns
are `mirrored' with respect to the central line, $x=1/2$.}
\end{figure}

\begin{example}
\label{Sawtooth}Figure 1 depicts the graphs of the identity (main diagonal),
the shift (or sawtooth) map $S_{2}:x\mapsto 2x(\func{mod}1)$, its second
iterate, $S_{2}^{2}$, and its third iterate, $S_{2}^{3}$. The vertical,
dashed lines rise at the endpoints of the intervals $P_{\pi }\neq \emptyset $
of points $x$ defining the allowed patterns $\pi \in \mathcal{S}_{4}$. We
conclude that $S_{2}$ has $18$ allowed $4$-patterns and hence $6$ forbidden $%
4$-patterns, namely:%
\begin{equation}
\left\langle 0,2,3,1\right\rangle ,\left\langle 1,0,2,3\right\rangle
,\left\langle 1,3,2,0\right\rangle ,\left\langle 2,0,1,3\right\rangle
,\left\langle 3,1,0,2\right\rangle ,\left\langle 3,2,0,1\right\rangle .
\label{S2}
\end{equation}%
We leave as an exercise to check that $S_{2}$ has no forbidden patterns of
length $3$.
\end{example}

We turn next to the question of finding natural conditions on $f$ that allow
the existence of forbidden patterns of any length. Given $\varepsilon >0$
arbitrarily small, there exists by Definition \ref{Top(f)} a product
partition $\iota _{0}$ such that
\[
\left\vert h_{top}^{\ast }(f)-h_{top}^{\ast }(\mathbf{S}^{\iota
})\right\vert <\frac{\varepsilon }{2}
\]%
whenever $\left\Vert \iota \right\Vert \leq \left\Vert \iota _{0}\right\Vert
$. Furthermore, there exists by Definition \ref{Top1and2} a length $L_{0}$
such that
\[
\left\vert h_{top}^{\ast }(\mathbf{S}^{\iota })-\frac{1}{L}\log N(\mathbf{S}%
^{\iota },L)\right\vert <\frac{\varepsilon }{2}
\]%
whenever $L\geq L_{0}$, where $N(\mathbf{S}^{\iota },L)$\textbf{\ }is the
number of \textit{allowed} ordinal $L$-patterns of the symbolic dynamic $%
\mathbf{S}^{\iota }$ with respect to the coarse-graining $\iota $.
Therefore, chosen $\iota $ sufficiently fine and $L$ sufficiently large, we
have
\[
\left\vert h_{top}^{\ast }(f)-\frac{1}{L}\log N(\mathbf{S}^{\iota
},L)\right\vert <\varepsilon ,
\]%
hence,%
\begin{equation}
N(\mathbf{S}^{\iota },L)=e^{Lh_{top}^{\ast }(f)}+\mathcal{O}_{L}(\varepsilon
),  \label{N(R,L)}
\end{equation}%
where the term $\mathcal{O}_{L}(\varepsilon )$ depends also on $L$, as
indicated by the subindex, and $\mathcal{O}_{L}(\varepsilon )\rightarrow 0$
when $\varepsilon \rightarrow 0$ (or $\left\Vert \iota \right\Vert
\rightarrow 0$). Since according to Stirling's formula,%
\[
L!\varpropto \left( L/e\right) ^{L}\sqrt{2\pi L}=e^{L(\log L-1)+(1/2)\log
2\pi L}
\]%
(where $\varpropto $ stands for \textquotedblleft
asymptotically\textquotedblright ), the number of \textit{possible} ordinal $%
L$-patterns, $\left\vert \mathcal{S}_{L}\right\vert =L!$, grows
superexponentially with $L$, we conclude from (\ref{N(R,L)}) that the
symbolic dynamic $\mathbf{S}^{\iota }$ has forbidden patterns whenever $%
h_{top}^{\ast }(f)$ is finite. Intuitively speaking, the same must happen
with maps whose dynamic can be well approximated by simple observations.

\begin{theorem}
\label{main2}Let $I\subset \mathbb{R}^{q}$ be a finite interval and $%
f:I\rightarrow I$ a continuous map. Then%
\[
\lim_{\left\Vert \iota \right\Vert \rightarrow 0}N(\mathbf{S}^{\iota
},L)=N(f,L),
\]%
where $N(f,L)$ is the number of allowed patterns of $f$ with length $L$.
\end{theorem}

\begin{pf}
Fix $L\geq 2$ and suppose that there exists an ordinal pattern $\pi
(x)=\left\langle \pi _{0},...,\pi _{L-1}\right\rangle $ defined by $x\in I$
under $f$ that is not visible to any symbolic dynamic $\mathbf{S}^{\iota }$
with respect to a product partition $\iota $. Since ordinal patterns are
defined by inequalities (see (\ref{patt})), there exists by continuity $\eta
>0$ such that $\left\Vert x-y\right\Vert <\eta $ implies $\pi (x)=\pi (y)$,
the ordinal pattern defined by $y$. This means that, in the limiting process
$\left\Vert \iota \right\Vert $ $\rightarrow 0$, we do take account of all
ordinal $L$-patterns that the orbits of $f$ can produce.
\end{pf}

From (\ref{N(R,L)}) it follows then,

\begin{cor}
\label{Corallary}Under the hypotheses of Theorem \ref{main2}, the number of
allowed $L$-patterns of $f$ grows with $L$ as%
\begin{equation}
\left\vert \{P_{\pi }\neq \varnothing :\pi \in \mathcal{S}_{L}\}\right\vert
\varpropto e^{Lh_{top}^{\ast }(f)},  \label{growth}
\end{equation}%
provided $h_{top}^{\ast }(f)<\infty $.
\end{cor}

Since checking the technical condition $h_{top}^{\ast }(f)<\infty $ is in
practice more difficult than checking directly the growth rate of the
allowed patterns of $f$, we use now Theorem \ref{main} to provide more
natural conditions for (\ref{growth}).

\begin{cor}
If $I\subset \mathbb{R}^{q}$ is a finite interval and $f:I\rightarrow I$ an
expansive map with $h_{top}(f)<\infty $, then (\ref{growth}) holds true.
\end{cor}

The bottom line is that interval maps have forbidden patterns under quite
general conditions and that they proliferate superexponentially with the
length $L$ as%
\[
\left\vert \{P_{\pi }=\varnothing :\pi \in \mathcal{S}_{L}\}\right\vert
\varpropto L!-e^{Lh_{top}^{\ast }(f)}
\]%
(more on this in Sect. 4.2).

Apart from the superexponential scaling law with $L$, it is quite difficult
to make more specific statements on the forbidden patterns of a given map $f$
like, for instance, the minimal length of its forbidden patterns. One
important exception is the shift transformation on sequence spaces.

\begin{example}
\label{forbshift}As in (\ref{sequence}), let $A^{\mathbb{N}_{0}}$ be the
one-sided sequence space of the symbols $A=\{a_{1},a_{2},...,a_{\left\vert
A\right\vert }\}$, where $a_{1}<a_{2}<...<a_{\left\vert A\right\vert }$, $A^{%
\mathbb{N}_{0}}$ being endowed with the lexicographic order $\preceq $
defined for $\mathbb{\alpha }\neq \mathbb{\alpha }^{\prime }$ as$\;$
\begin{equation}
\mathbb{\alpha }\prec \mathbb{\alpha }^{\prime }\;\Leftrightarrow \left\{
\begin{array}{cc}
\;\mathbb{\alpha }_{i}=\mathbb{\alpha }_{i}^{\prime } & \mbox{for}\text{ }%
0\leq i\leq n \\
\text{\mbox{and}} &  \\
\mathbb{\alpha }_{n+1}<\mathbb{\alpha }_{n+1}^{\prime } &
\end{array}%
\right. .  \label{lexico}
\end{equation}%
Furthermore, let $\Sigma :A^{\mathbb{N}_{0}}\rightarrow A^{\mathbb{N}_{0}}$
be the corresponding \textit{shift }%
\begin{equation}
(\mathbb{\alpha }_{0},\mathbb{\alpha }_{1},\mathbb{\alpha }_{2},...)\mapsto (%
\mathbb{\alpha }_{1},\mathbb{\alpha }_{2},\mathbb{\alpha }_{3},...).
\label{one-sided shift}
\end{equation}%
Then one can prove \cite{Amigo4}:

\begin{enumerate}
\item For every $L\leq \left\vert A\right\vert +1$, $\Sigma $ has no
forbidden patterns.

\item For every $L\geq \left\vert A\right\vert +2$, $\Sigma $ has forbidden
root patterns of length $L$. For instance if $\left\vert A\right\vert $ is
even, then the ordinal patterns of length $L=\left\vert A\right\vert +2$
\begin{eqnarray*}
&&\left\langle 0,2,...,\left\vert A\right\vert ,\left\vert A\right\vert
+1,...,3,1\right\rangle , \\
&&\left\langle \left\vert A\right\vert +1,\left\vert A\right\vert
-1,...,1,0,2,...,\left\vert A\right\vert -2,\left\vert A\right\vert
\right\rangle
\end{eqnarray*}%
and
\[
\left\langle \left\vert A\right\vert -1,\left\vert A\right\vert
-3,...,1,0,2,...,\left\vert A\right\vert ,\left\vert A\right\vert
+1\right\rangle
\]%
are forbidden. Moreover, if $\pi =\left\langle \pi _{0},\pi _{1},...,\pi
_{L-2},\pi _{L-1}\right\rangle $ is forbidden for $\Sigma $, then its \emph{%
mirrored pattern}%
\[
\pi _{\text{\emph{mirrored}}}=\left\langle \pi _{L-1},\pi _{L-2},...,\pi
_{1},\pi _{0}\right\rangle
\]%
is also forbidden for $\Sigma $.
\end{enumerate}
\end{example}

\section{Properties of the forbidden patterns}

To complete the picture, we briefly review in this section the three more
important properties of forbidden patterns (see also \cite{Amigo5,Amigo4}).

\subsection{Invariance under order-isomorphims}

Since ordinal patterns are not directly related to measure-theoretical
properties, isomorphic (or conjugate) dynamical systems need not have the
same forbidden patterns, unless the isomorphism between them preserves not
only measure but also linear order (supposing both state spaces are linearly
ordered).

For instance, the graphical technique used in Figure 1 reveals that the
logistic map $f(x)=4x(1-x)$, $0\leq x\leq 1$, has the forbidden $3$-pattern $%
\left\langle 2,1,0\right\rangle $ \cite{Amigo5}, i.e., there are no three
consecutive points in any orbit generated by the logistic map, forming a
strictly decreasing trio. However, it follows from the general results
stated in Example \ref{forbshift} that the one-sided $(\frac{1}{2},\frac{1}{2%
})$-Bernoulli shift \cite{Katok} has no forbidden patterns of length $3$,
despite being conjugate to the logistic map (the interval $[0,1]$ endowed
with the measure $\frac{dx}{\pi \sqrt{x(1-x)}}$). The reason is that the
corresponding isomorphism, actually the symbolic dynamic $\mathbf{S}^{\iota
}:[0,1]\rightarrow \{0,1\}^{\mathbb{N}_{0}}$ with respect to the generating
partition $\iota =\{I_{0}=[0,\frac{1}{2}),I_{1}=[\frac{1}{2},1]\}$, is not
order-preserving: e.g.,%
\[
\mathbf{S}^{\iota }(\tfrac{1}{4})=(0,\bar{1})<\mathbf{S}^{\iota }(\tfrac{3}{4%
})=(\bar{1}),
\]%
where the overbar denotes indefinite repetition of the binary digit and
binary strings are ordered lexicographically, while%
\[
\mathbf{S}^{\iota }(\tfrac{1}{2})=(1,1,\bar{0})>\mathbf{S}^{\iota }(1)=(1,%
\bar{0}).
\]

\begin{definition}
Given two linearly ordered sets $(X_{1},\leq _{1})$ and $(X_{2},\leq _{2})$,
two maps $T_{1}:X_{1}\rightarrow X_{1}$ and $T_{2}:X_{2}\rightarrow X_{2}$
and an invertible map $\Phi :X_{1}\rightarrow X_{2}$ such that $\Phi \circ
T_{1}=T_{2}\circ \Phi $, we say that $T_{1}$ and $T_{2}$ are \textit{%
order-isomorphic} if $\Phi $ is order-preserving (i.e., $x\leq _{1}y$
implies $\Phi (x)\leq _{2}\Phi (y)$).
\end{definition}

It is trivial that order-isomorphic maps have the same allowed (and, hence,
forbidden) ordinal patterns. Let us see next an interesting example of an
order-isomorphism.

\begin{example}
\label{coding}Let $A=\{0,1,...,b-1\}$ and consider the two-sided sequence
space with alphabet $A$,%
\[
A^{\mathbb{Z}}=\{\mathbb{\alpha }=(\mathbb{\alpha }_{n})_{n\in \mathbb{Z}%
}=(...,\mathbb{\alpha }_{-1},\mathbb{\alpha }_{0},\mathbb{\alpha }_{1},...):%
\mathbb{\alpha }_{n}\in A\}.
\]%
With the notation $\mathbb{\alpha }_{-}$ for the `\textit{left sequence'} $(%
\mathbb{\alpha }_{-n})_{n\in \mathbb{N}}$ of the `bisequence' $\mathbb{%
\alpha }\in A^{\mathbb{Z}}$ and $\mathbb{\alpha }_{+}$ for its `\textit{%
right sequence'} $(\mathbb{\alpha }_{n})_{n\in \mathbb{N}_{0}}$, we define a
linear order $\preceq $ in $A^{\mathbb{Z}}$ by%
\begin{equation}
\mathbb{\alpha }\prec \mathbb{\alpha }^{\prime }\;\Leftrightarrow \left\{
\begin{array}{l}
\;\mathbb{\alpha }_{+}<\mathbb{\alpha }_{+}^{\prime } \\
\text{\mbox{or}} \\
\mathbb{\alpha }_{-}<\mathbb{\alpha }_{-}^{\prime }\text{ \mbox{if} }\mathbb{%
\alpha }_{+}=\mathbb{\alpha }_{+}^{\prime }%
\end{array}%
\right. \text{ ,}  \label{lex}
\end{equation}%
where $\leq $ between right (resp. left) sequences denotes lexicographical
order in $A^{\mathbb{N}_{0}}$ (resp. $A^{\mathbb{N}}$), see (\ref{lexico}).
If $\mathcal{N}$ denotes the null set of bisequences $\mathbb{\alpha }$
eventually terminating in an infinite string of $(b-1)$s in either
direction, then the map $\psi :A^{\mathbb{Z}}\backslash \mathcal{N}%
\rightarrow $ $[0,1)\times \lbrack 0,1)\equiv \lbrack 0,1)^{2}$ (here
\textquotedblleft\ $\backslash $\textquotedblright\ stands for set
difference) defined by
\begin{equation}
\psi :(\mathbb{\alpha }_{-},\mathbb{\alpha }_{+})\mapsto \left(
\sum_{n=1}^{\infty }\mathbb{\alpha }_{-n}b^{-n},\sum_{n=0}^{\infty }\mathbb{%
\alpha }_{n}b^{-(n+1)}\right) ,  \label{psibar}
\end{equation}%
is one-to-one and order-preserving. As a matter of fact, the order (\ref{lex}%
) in $A^{\mathbb{Z}}$ corresponds via $\psi $ to the lexicographical order
in $[0,1)^{2}$, so we may call $\preceq $ the lexicographical order in $A^{%
\mathbb{Z}}$. In sum, $\psi $ is an order-isomorphism, both $A^{\mathbb{Z}%
}\backslash \mathcal{N}$ and $[0,1)^{2}$ being endowed with the
lexicographic order.
\end{example}

As way of application, consider the (non-continuous!) \textit{baker map}, $%
B:[0,1)^{2}\rightarrow \lbrack 0,1)^{2}$, where
\[
B(x,y)=\left\{
\begin{array}{lc}
(2x,\frac{1}{2}y), & 0\leq x<\frac{1}{2}, \\
(2x-1,\frac{1}{2}y+\frac{1}{2}), & \frac{1}{2}\leq x<1.%
\end{array}%
\right.
\]%
If now $\overline{\Sigma }$ denotes a two-sided shift on two-symbol
sequences, then $B$ and $\overline{\Sigma }$ are order-isomorphic\footnote{%
They are even conjugate as dynamical systems if $\overline{\Sigma }$ is the
two-sided $(\frac{1}{2},\frac{1}{2})$-Bernoulli shift and $[0,1)^{2}$ is
endowed with Lebesgue measure.}, modulo the null set $\mathcal{N}$, via the
`coding map' $\psi :A^{\mathbb{Z}}\backslash \mathcal{N}\rightarrow $ $%
[0,1)^{2}$ given in (\ref{psibar}). It follows that the baker map and the
two-sided shift on two-symbol sequences have the same allowed and forbidden
patterns.

Even more is true. First of all, one- and two-sided shifts on (bi-)sequences
ordered lexicographically (see (\ref{lexico}) and (\ref{lex}), respectively)
can be proven to have the same forbidden patterns \cite{Amigo4}.
Furthermore, one can also prove along the same lines as in Example \ref%
{coding} that the sawtooth map $x\longmapsto 2x$ (\mbox{mod 1}) and the
one-sided shift (\ref{one-sided shift}) on two-symbol sequences are
order-isomorphic (modulo the null set of sequences terminating in $\bar{1}$%
). We conclude that the allowed and forbidden $4$-patterns of the baker map
are precisely those exhibited in Figure 1 and listed in (\ref{S2}),
respectively.

\subsection{Growth with length: outgrowth patterns}

According to Corollary \ref{Corallary}, for every continuous self map $f$ on
a finite $q$-dimensional interval with finite topological permutation
entropy, there exists $\pi \in \mathcal{S}_{L}$, $L\geq 2$, which cannot
occur in any orbit. Moreover, if $\pi =\left\langle \pi _{0},...,\pi
_{L-1}\right\rangle $ is forbidden for $f$, then it is easy to see that the $%
2(L+1)$ patterns of length $L+1$,
\begin{eqnarray*}
&&\left\langle L,\pi _{0},...,\pi _{L-1}\right\rangle ,\left\langle \pi
_{0},L,\pi _{1},...,\pi _{L-1}\right\rangle ,...,\left\langle \pi
_{0},...,\pi _{L-1},L\right\rangle , \\
&&\left\langle 0,\pi _{0}+1,...,\pi _{L-1}+1\right\rangle ,\left\langle \pi
_{0}+1,0,\pi _{1}+1,...,\pi _{L-1}+1\right\rangle ,...,\left\langle \pi
_{0}+1,...,\pi _{L-1}+1,0\right\rangle ,
\end{eqnarray*}%
are also forbidden for $f$. A weak form of the converse holds also true: if $%
\left\langle L,\pi _{0},...,\pi _{L-1}\right\rangle $, $\left\langle \pi
_{0},L,...,\pi _{L-1}\right\rangle $, $...$, $\left\langle \pi _{0},...,\pi
_{L_{0}-1},L\right\rangle \in \mathcal{S}_{L+1}$ are forbidden, then $%
\left\langle \pi _{0},...,\pi _{L-1}\right\rangle \in \mathcal{S}_{L}$ is
also forbidden.

In turn, each of these forbidden patterns of length $L+1$ belonging to the
`first generation',\ will generate a `second generation' of forbidden
patterns of length $L+2$, not necessarily all different, etc.. Observe that
all these forbidden patterns generated by $\pi $ in the $n$th generation
have the form
\begin{equation}
\left\langle \ast ,\pi _{0}+n,\ast ,\pi _{1}+n,\ast ,...,\ast ,\pi
_{L-1}+n,\ast \right\rangle \in \mathcal{S}_{N}  \label{piele}
\end{equation}%
(the wildcard $\ast $ stands eventually for any other entries of the
pattern), with $n=0,1,...,N-L$, where $N-L\geq 1$ is the number of wildcards
$\ast \in \{0,1,...,n-1,L+n,...,N-1\}$ (with $\ast \in \{L,...,N-1\}$ if $n=0
$ and $\ast \in \{0,...,N-L-1\}$ if $n=N-L$). Forbidden patterns of the form
(\ref{piele}), where $\pi =\left\langle \pi _{0},...,\pi _{L-1}\right\rangle
$ is forbidden, are called \textit{outgrowth forbidden patterns}. If $%
\mathcal{S}_{N}^{out}(\pi )$ denotes the set of outgrowth forbidden $N$%
-patterns of $\pi $, the it can be proven \cite{Amigo4} that there exist
constants $0<c,d<1$ such that%
\begin{equation}
(1-d^{N})N!<\left\vert \mathcal{S}_{N}^{out}(\pi )\right\vert <(1-c^{N})N!.
\label{outgrowth}
\end{equation}

Forbidden patterns that are not outgrowth patterns of other forbidden
patterns of shorter length are called \textit{forbidden root patterns} since
they can be viewed as the root of the tree of forbidden patterns spanned by
the outgrowth patterns they generate, branching taking place when going from
one length (or generation) to the next. Thus (\ref{outgrowth}) shows that
alone the number of outgrowth $N$-patterns of a given forbidden $L$-pattern,
$N>L$, follows a superexponential growth law with $N$.

\subsection{\protect\medskip Robustness against noise}

Finally, let us elaborate on the persistence of forbidden patterns when the
observed data are distorted by small perturbations, a property we refer to
as robustness against observational noise. As already mentioned in the
Introduction, forbidden patterns are robust against observational noise on
account of being defined by inequalities. Were not for this property,
forbidden patterns would not be useful in applications.

The sort of applications we have in mind belong to the detection of
determinism in univariate and multivariate time series analysis, since
random real-valued time series have no forbidden patterns with probability $1
$; see \cite{Amigo5}\ for the intricacies of the scalar case when the
sequences are finite and contaminated with (additive) white noise, i.e.,
when the time series have the form
\begin{equation}
z_{k}=f^{k}(x)+\eta _{k},  \label{noisy}
\end{equation}%
with $f$ being a one-dimensional interval map and $\eta _{k}$ real-valued
(and properly bounded), independent, equally distributed (i.i.d.) random
variables. A $2$-dimensional time series contaminated with white noise will
be considered in the next section. The case of colored noise (i.e., random
variables with correlation) is more difficult and is currently under
investigation; ias a matter of fact, numerical sequences of the form (\ref%
{noisy}) are often used to generate colored noise.

\section{Numerical simulations}

We demonstrate numerical evidence for the existence of forbidden ordinal
patterns in multi-dimensional maps. Of course, direct simulation of
dynamical systems directly yields only \emph{allowed} order patterns. The
failure to observe any given order pattern/permutation in any finite time
series does not mean of course that it is forbidden (probability zero) but
only that its probability is sufficiently low in the measure induced by the
natural dynamics that it has not yet been seen.

However, with sufficiently reasonable $L$ (as effort and memory increases
radically with $L$) and robust computational ability we can infer in many
cases, robustness of forbidden patterns by examining the convergence of
allowed patterns with $N$, the number of order patterns (of length $L$)
emitted by the data generating source. In particular, we suggest examining
the logarithmic ratio of the cardinality of all patterns to the number of
observed patterns $\log \left( L!/P_{\text{obs}}\right) $ vs $\log N$. If a
system has a \textquotedblleft core\textquotedblright\ of forbidden
patterns, as with deterministic systems, then we expect that this ratio will
decline with $N$ and eventually level off with increasing $N$, assuming the
asymptotic behavior can be observed. Here, $P_{\text{obs}}$ is the naive,
biased-downward, estimator of the unknown $P_{\text{allowed}}$.

When $N$ is much larger than $P_{\text{allowed}}$, $P_{\text{obs}}$ is
likely to be a good estimator, assuming most patterns have a reasonable
probability of occurring. With increasing $L$, however, this is difficult to
achieve practically because of memory limitations, as the identities and
counts of each observed patterns (a subset of the allowed patterns) must be
retained. The number of allowed patterns increases exponentially with $L$ in
deterministic chaos, and faster than exponentially with noise, and therefore
one must increase $N$, the number of iterates, substantially to permit a
commensurately large number of distinct patterns to be actually observed.

This motivates using a superior statistical estimator of $P_{\text{allowed}}$%
. This equivalent problem has a significant history, motivated especially
from the ecology community. Consider a situation where one can observe a
finite sample of individual organisms, from a presumably large population.
What is the estimated number of distinct species, the biodiversity, and how
can we estimate this given the individual counts of observed species? (For
reviews of approaches to this problem see~\cite{Review1,Review2}.) This is
analogous to our situation where we can distinguish individual order
patterns but each observation is drawn from the natural distribution induced
by typical orbits of the dynamical system. For our needs we wish to go
reasonably deep into the undersampled regime and impose few probabilistic
priors. We adopt the non-parametric estimator of Chao~\cite{Chao84},
motivated by comments in the reviews and our experience, as a simple but
reasonably effective improvement:
\begin{equation}
P_{\text{Chao}}=P_{\text{obs}}+\frac{c_{1}^{2}}{2c_{2}^{2}},
\end{equation}%
where $c_{k}$ are the \textquotedblleft meta-counts\textquotedblright\ of
observations, i.e. $c_{1}$ is the number of distinct ordinal patterns which
were observed exactly once in the sample, $c_{2}$ the number which were
observed exactly twice, etc. In practice this is accomplished by counting
frequencies of observed patterns through a hash table, and in a second
phase, counting the frequencies of such frequencies with a similar hash
table. Note that if the sample size is particularly small (relative to what
is necessary to see a substantial fraction of allowed patterns), $P_{\text{%
Chao}}$ will still be an underestimate. Consider that its maximum value is
obtained with $c_{1}=N-1$ and $c_{2}=1$, i.e. one doubleton and all
remaining observations being unique (all unique naturally leads to an
undefined estimate), and so $P_{\text{Chao}}$ is bounded by $(P_{\text{obs}%
}^{2}+1)/2$. Bunge and Fitzpatrick~\cite{Review1} call $P_{\text{Chao}}$ to
be an \textquotedblleft estimated lower bound\textquotedblright . We believe
that no statistical estimator can perform well in the extremely undersampled
regime and there is no substitute for substantial computational effort when $%
L$ becomes sufficiently large; however, we will see an improvement over the
naive estimator.

\begin{figure}[tbp]
\par
\centerline{\includegraphics[width=0.7%
\columnwidth]{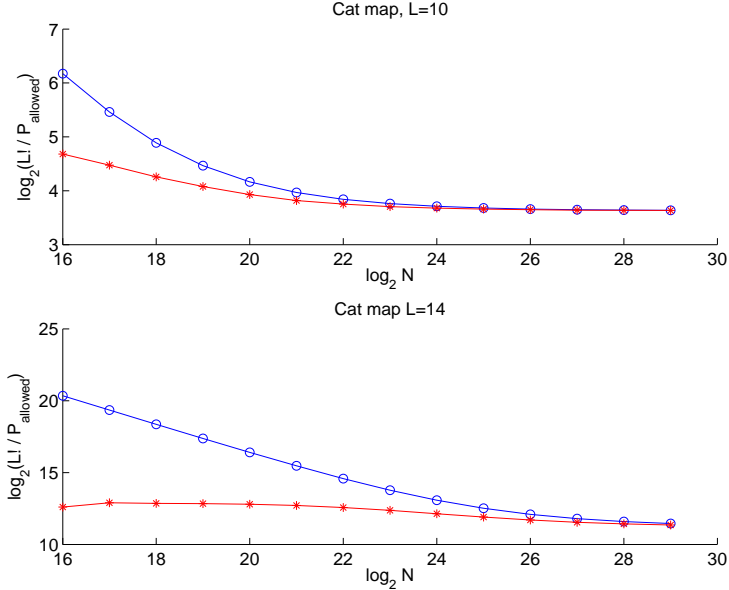}} \label{fig:catmap1}
\caption{Convergence of estimated forbidden patterns with $N$, cat
map. Blue circles (o) are for $P_\text{allowed} = P_\text{obs}$, red
asterisks (*) have $P_\text{allowed} = P_\text{Chao}$. Top, $L=10$,
bottom $L=14$. Both figures show clear evidence of convergence to a
constant, evidence of true forbidden patterns as $N \rightarrow
\infty$. In the lower figure especially, the improved estimator
$P_\text{Chao}$ ``senses'' the approach to a convergence earlier
than the naive counting estimator. Note the differing scales on the
$y$-axes.}
\end{figure}

\begin{figure}[tbp]
\centerline{\includegraphics[width=0.7%
\columnwidth]{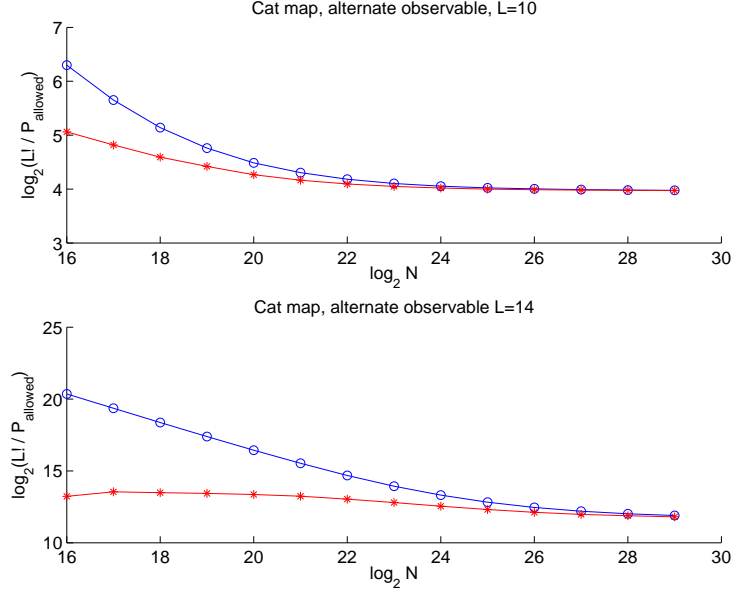}} \label{fig:catmap2}
\caption{Convergence of estimated forbidden patterns with $N$, cat
map,
alternative observable. Blue circles (o) are for $P_\text{allowed} = P_\text{%
obs}$, red asterisks (*) have $P_\text{allowed} = P_\text{Chao}$. Top, $L=10$%
, bottom $L=14$. Both figures show clear evidence of convergence to a
constant, evidence of true forbidden patterns as $N \rightarrow \infty$. In
the lower figure especially, the improved estimator $P_\text{Chao}$
``senses'' the approach to a convergence earlier than the naive counting
estimator. Note the differing scales on the $y$-axes.}
\end{figure}

Our first numerical example is Arnold's \textquotedblleft
cat\textquotedblright\ map: $(x,y)\rightarrow (x+y\mod 1,x+2y\mod 1)$. We
start with initial conditions drawn uniformly in $[0,1)\times \lbrack 0,1)$,
and iterate. Ordinal patterns are computed using order relations on the $x$%
-coordinate only. Figure~2 shows the strong numerical evidence for forbidden
patterns characteristic of deterministic systems. As a demonstration of the
genericity of the results, Figure~3 shows the equivalent except that the
observable upon which order patterns were computed is $3x^{3}-y$. Results
are nearly identical, as one expects.

\begin{figure}[tbp]
\centerline{\includegraphics[width=0.7%
\columnwidth]{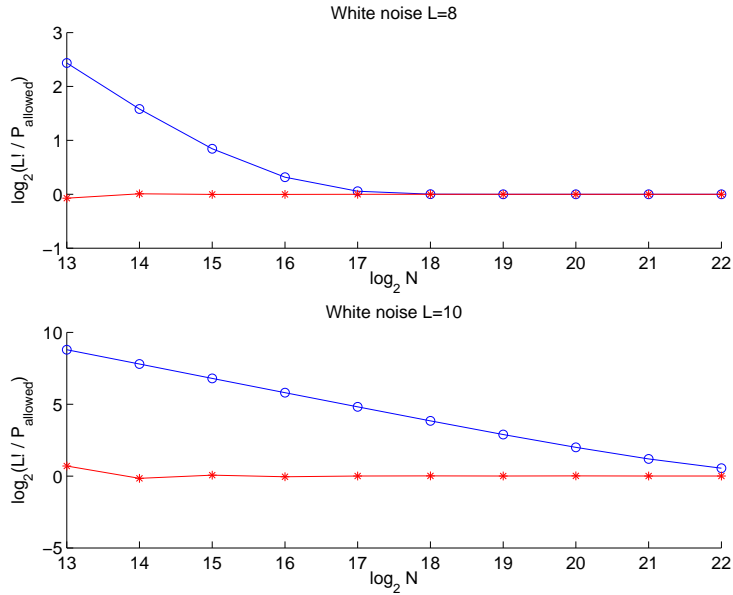}} \label{fig:noise}
\caption{Convergence of estimated forbidden patterns with $N$,
i.i.d. noise. Blue circles (o) are for $P_\text{allowed} =
P_\text{obs}$, red asterisks
(*) have $P_\text{allowed} = P_\text{Chao}$. Top, $L=8$, bottom $L=10$. $P_%
\text{obs}$ shows convergence to zero forbidden pattterns; $P_\text{Chao}$
estimates zero forbidden patterns well before convergence of naive
estimator. }
\end{figure}

By comparison, consider Figure~4, generated by an i.i.d. noise source
(ordinal patterns are insensitive to changes in distribution). Here, the
observed patterns imply convergence to zero forbidden patterns with
increasing $N$. More remarkably the estimator $P_{\text{Chao}}$ senses this
long before and predicts zero forbidden patterns with orders of magnitude
lower $N$, apparently because the assumptions made by the estimator of
equiprobable patterns for both observed and unobserved are exactly fulfilled.

\begin{figure}[tbp]
\centerline{\includegraphics[width=0.7%
\columnwidth]{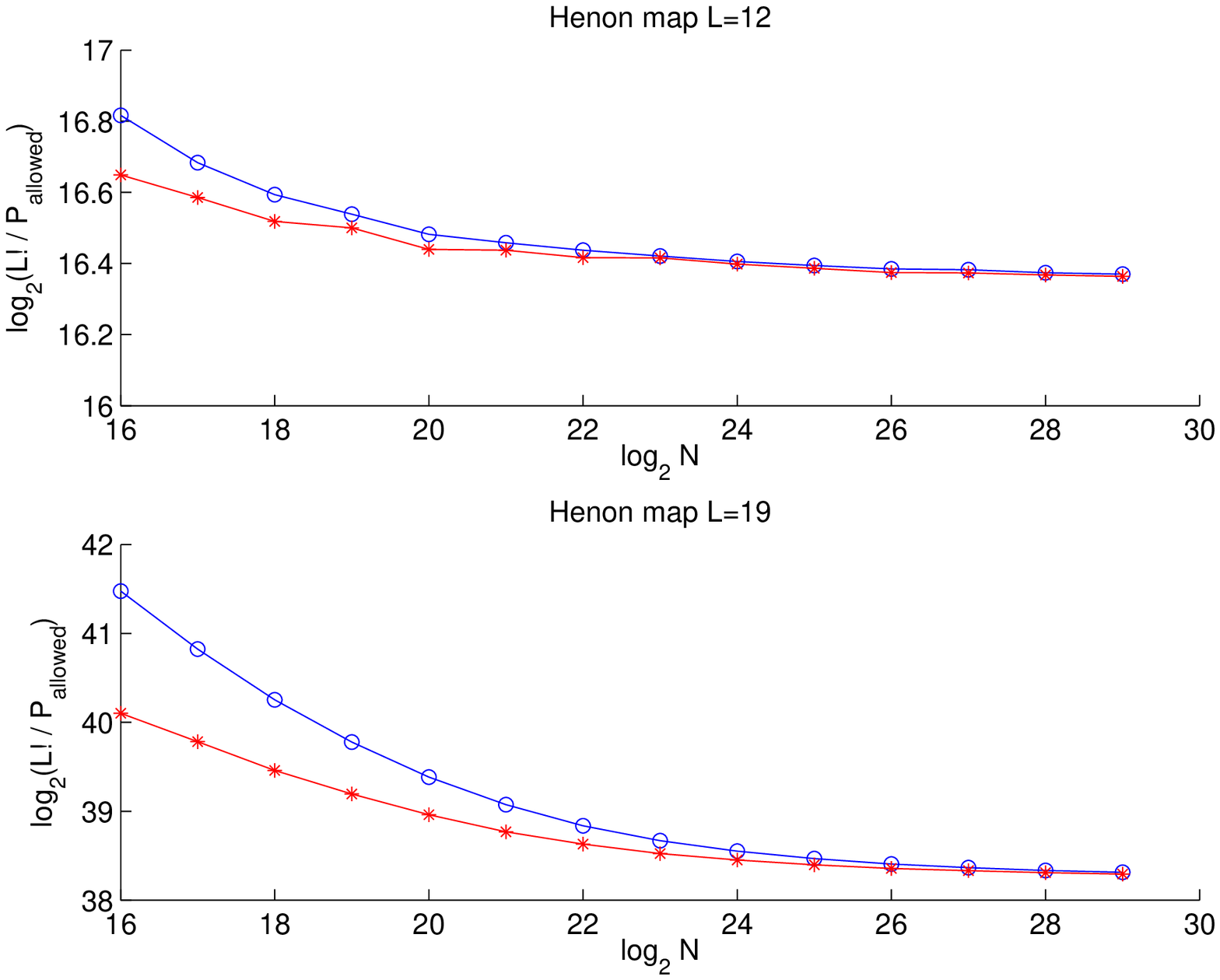}} \label{fig:henon}
\caption{Convergence of estimated forbidden patterns with $N$,
H\'{e}non map. Blue circles (o) are for $P_\text{allowed} =
P_\text{obs}$, red
asterisks (*) have $P_\text{allowed} = P_\text{Chao}$. Top, $L=12$, bottom $%
L=19$. Both naive and improved estimators show convergence to a finite
number of forbidden patterns with increasing $N$. Note scale of $y$-axes.}
\end{figure}

\begin{figure}[tbp]
\centerline{\includegraphics[width=0.7%
\columnwidth]{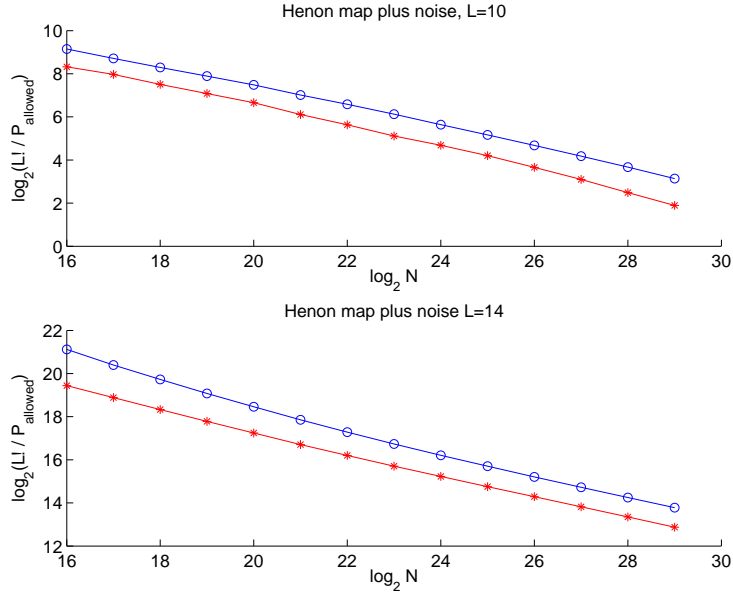}} \label{fig:henonnoise}
\caption{Lack of cnvergence of estimated forbidden patterns with $N$, H\'{e}%
non map with additive i.i.d. noise. Blue circles (o) are for $P_\text{allowed%
} = P_\text{obs}$, red asterisks (*) have $P_\text{allowed} = P_\text{Chao}$%
. Top, $L=12$, bottom $L=19$. Both naive and improved estimators show
continued increase in allowed patterns (decrease in forbidden patterns) with
increasing $N$.}
\end{figure}

The cat map may be seen as too \textquotedblleft easy\textquotedblright\ and
so we turn to a chaotic system, the H\'{e}non map, $(x,y)\rightarrow
(1-ax^{2},bx),\,a=1.4,b=0.3$, observable being the $x$-coordinate. This map
is not uniformly hyperbolic, more characteristic of real dynamics seen in
nature. In Figure~5, we see convergence to a finite core of forbidden
patterns with larger $N$. Note that the performance of $P_{\text{Chao}}$ is
still improved over the naive estimator but it is not as good as with noise,
because with real dynamics there is a wide variation in the probability of
the various allowed patterns, and so larger $N$ feels the 'tail' of the
distribution of rare patterns. By comparison consider Figure~6, which shows
results from the same dynamics but each observable was contaminated with
uniform i.i.d noise $\eta \in \lbrack 0,0.2)$. This time, increasing $N$
clearly shows increasing allowed/decreasing forbidden patterns, proportional
to $N$ as expected with noise. The behavior with $N$ cleanly distinguishes
low-dimensional dynamics from noise.

As a philosophical point it is true that the ``noise'' generator in a
computer software is but a deterministic dynamical system on its own, but in
practice it has an extremely long period and virtually no correlation, and
hence if one wanted to see order pattern scaling different from true noise,
one would need exceptionally long $L$ and impractically astronomical memory
requirements. We use a validated high-quality random number generator~\cite%
{mersennetwister} from the Boost C++ library.

\section{\noindent Conclusion}

We showed that $n$-dimensional interval maps have forbidden ordinal patterns
under the following two sufficient conditions: (i) continuity, and (ii)
finite topological permutation entropy (Corollary 3). The second condition,
that can be difficult to check, may be replaced by (ii') finite topological
entropy if the first condition is replaced by (i') expansiveness (Corollary
4). In any case, we conjecture as a working hypothesis that the existence of
forbidden patterns is a general feature of the interval maps encountered in
practice.

Interestingly enough, the existence of forbidden patterns can be used as an
indicator of determinism in univariate and multivariate time series
analysis, since sequences generated by unconstrained random processes taking
values on intervals have no forbidden patterns with probability one. The
application of these ideas requires some care since real sequences are
\textit{finite} (making possible that random sequences have `false'
forbidden patterns with finite probability) and \textit{noisy} (blurring the
difference between determinism and randomness). The numerical simulations of
Sect. 5 have provided ample evidence of all these issues, in particular of
the robustness of forbidden patterns against observational white noise. In
so doing we have also used Chao's estimator for the number of classes in a
population.

\section*{Acknowledgments}

This work has been financially supported by the Spanish Ministry of
Education and Science, grant MTM2005-04948 and European FEDER Funds.

\end{document}